\documentclass[amsmath,amssymb,twocolumn,superscriptaddress,floatfix]{revtex4-1}

\usepackage{hyperref}
\usepackage{graphicx}
\usepackage{textcomp}
\usepackage{nameref}
\usepackage{xargs}   
\usepackage[pdftex,dvipsnames]{xcolor}
\usepackage{siunitx}
\usepackage{float}
\usepackage{lipsum}

\hypersetup{
    colorlinks,
    linkcolor={red!50!black},
    citecolor={blue!50!black},
    urlcolor={blue!80!black}
}

\newcommandx{\subfigcap}[1]{(#1)}

\newcommand{\SiV}{SiV$^-$}
\newcommand{\NV}{NV$^-$}

\begin{document}

\title{Single \SiV{} centers in low-strain nanodiamonds with bulk-like spectral properties and nano-manipulation capabilities}

\author{Lachlan J Rogers}
\thanks{These two authors contributed equally}
\affiliation{Department of Physics and Astronomy, Macquarie University, New South Wales 2109, Australia}
\affiliation{ARC Centre of Excellence for Engineered Quantum Systems (EQUS)}

\author{Ou Wang}
\thanks{These two authors contributed equally}
\affiliation{Institute for Quantum Optics, University Ulm, Albert-Einstein-Allee 11, D-89081 Ulm, Germany}
\affiliation{Center for Integrated Quantum Science and Technology (IQST), University Ulm, Albert-Einstein-Allee 11, D-89081 Ulm, Germany}

\author{Yan Liu}
\affiliation{Institute for Quantum Optics, University Ulm, Albert-Einstein-Allee 11, D-89081 Ulm, Germany}
\author{Lukas Antoniuk}
\affiliation{Institute for Quantum Optics, University Ulm, Albert-Einstein-Allee 11, D-89081 Ulm, Germany}
\author{Christian Osterkamp}
\affiliation{Institute for Quantum Optics, University Ulm, Albert-Einstein-Allee 11, D-89081 Ulm, Germany}
\affiliation{Center for Integrated Quantum Science and Technology (IQST), University Ulm, Albert-Einstein-Allee 11, D-89081 Ulm, Germany}
\affiliation{Department of Electron Devices and Circuits, University Ulm, Albert Einstein Allee 45, 89069 Ulm, Germany}
\author{Valery A. Davydov}
\affiliation{L.F.Vereshchagin Institute for High Pressure Physics, Russian Academy of Sciences, Troitsk, Moscow, 142190, Russia }
\author{Viatcheslav N. Agafonov}
\affiliation{GREMAN, UMR CNRS CEA 6157, Université F. Rabelais, F-37200 Tours, France}
\author{Andrea B. Filipovski}
\affiliation{Institute for Quantum Optics, University Ulm, Albert-Einstein-Allee 11, D-89081 Ulm, Germany}
\author{Fedor Jelezko}
\affiliation{Institute for Quantum Optics, University Ulm, Albert-Einstein-Allee 11, D-89081 Ulm, Germany}
\affiliation{Center for Integrated Quantum Science and Technology (IQST), University Ulm, Albert-Einstein-Allee 11, D-89081 Ulm, Germany}
\author{Alexander Kubanek}
\affiliation{Institute for Quantum Optics, University Ulm, Albert-Einstein-Allee 11, D-89081 Ulm, Germany}
\affiliation{Center for Integrated Quantum Science and Technology (IQST), University Ulm, Albert-Einstein-Allee 11, D-89081 Ulm, Germany}

\begin{abstract}
We report on the isolation of single \SiV{} centers in nanodiamonds. 
We observe the fine-structure of single \SiV{} center with improved inhomogeneous ensemble linewidth below the excited state splitting, stable optical transitions, good polarization contrast and excellent spectral stability under resonant excitation. 
Based on our experimental results we elaborate an analytical strain model where we extract the ratio between strain coefficients of excited and ground states as well the intrinsic zero-strain spin-orbit splittings. 
The observed strain values are as low as best values in low-strain bulk diamond. 
We achieve our results by means of H-plasma treatment of the diamond surface and in combination with resonant and off-resonant excitation. 
Our work paves the way for indistinguishable, single photon emission. 
Furthermore, we demonstrate controlled nano-manipulation via atomic force microscope cantilever of 1D- and 2D-alignments with a so-far unreached accuracy of about 10\,nm, as well as new tools including dipole rotation and cluster decomposition. 
Combined, our results show the potential to utilize \SiV{} centers in nanodiamonds for the controlled interfacing via optical coupling of individually well-isolated atoms for bottom-up assemblies of complex quantum systems. 
\end{abstract}

\maketitle

\section*{Introduction}

Experimental quantum optics has gradually shifted focus from testing fundamentals of quantum mechanics using isolated quantum systems such as atoms or photons \cite{haroche_controlling_2013,wineland_nobel_2013} towards the development of quantum technologies utilizing many connected quantum systems.
Applications include quantum computation (ion-traps and superconducting technology have achieved several quantum bits \cite{monroe_scaling_2013, devoret_superconducting_2013}) and low entropy systems, realized in atom-by-atom assemblies of up to $100$ atoms in one-dimensional defect-free arrays \cite{endres_atom-by-atom_2016}. 
The central challenge is to balance the need for quantum systems to be well isolated from their environment with the competing requirement that they interact with each other via well-controlled coupling.
Coupling mechanisms include magnetic interaction, dipole-dipole interaction, and long-range interaction between distant atoms mediated via photons as employed in photonic quantum technologies \cite{obrien_photonic_2009} or distributed quantum networks \cite{kimble_quantum_2008, ritter_elementary_2012}.
Efficient interfacing requires precise positioning on the relevant scale of the specific coupling mechanism, which is smaller than the electromagnetic field wavelength for optical coupling.

\begin{figure}
	\includegraphics[width=\columnwidth]{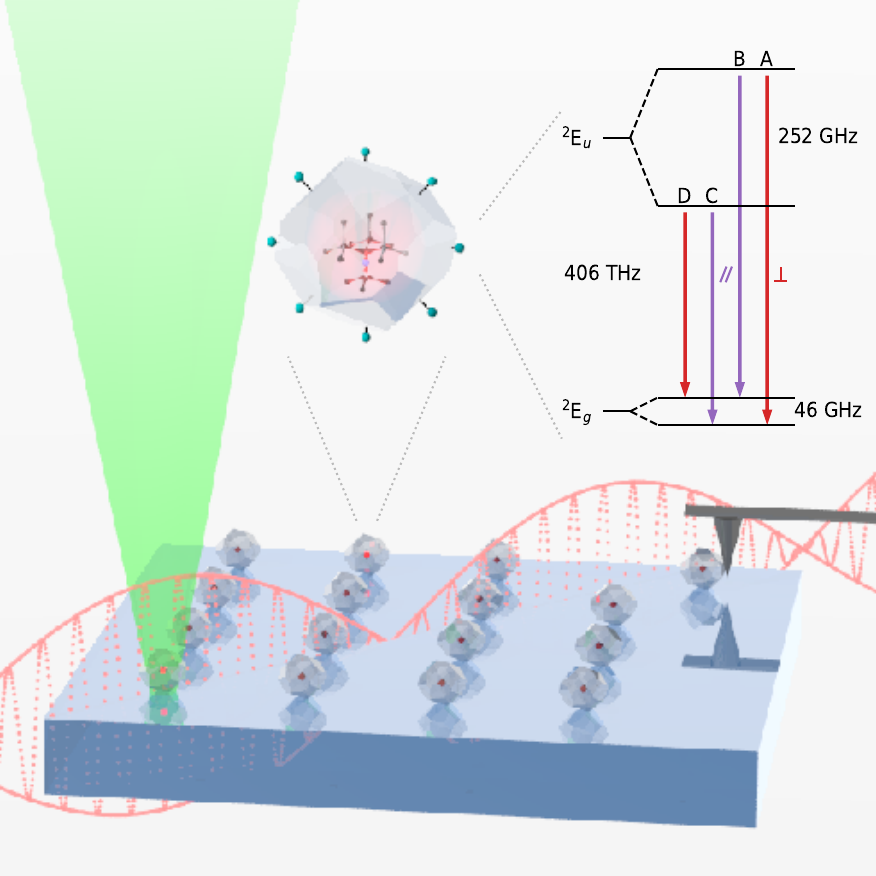}
	\caption{
		A conceptual design for single-emitter assemblies involves bottom-up formation 
		from pre-selected nanodiamonds.
		Hydrogen surface termination is found to improve the optical properties of \SiV{} centres in nanodiamonds, which could enable deterministic creation of emitter arrays for quantum applications.
		The optical properties of the \SiV{} centre in diamond arise from four optical transitions between doublet ground and excited states.
	}
	\label{fig:array-illustration}
\end{figure}

Semiconductor quantum optics is particularly well suited for constructing scalable devices satisfying this requirement, and the silicon-vacancy (\SiV{}) color center in diamond has emerged as an excellent optical-spin candidate.
Single \SiV{} centers have been coupled to all-diamond photonic platforms \cite{schroder_scalable_2017,sipahigil_integrated_2016} and are good sources of indistinguishable photons \cite{rogers_multiple_2014,sipahigil_indistinguishable_2014} with lifetime-limited spectral linewidth, low spectral diffusion in low-strain diamond hosts \cite{hepp_electronic_2014}, and a high Debye-Waller factor \cite{collins_annealing_1994}.
Arrays of bright \SiV{} centers have been created via ion implantation with a yield of up to 15 percent \cite{tamura_array_2014}. 
However a bottom-up approach utilizing single \SiV{} centers in preselected nanodiamonds (NDs) could enable  deterministic emitter assemblies where hardware errors could be corrected by replacing defective NDs (\autoref{fig:array-illustration}).
This offers a  wide-ranging flexibility for hybrid photonic platforms \cite{wolters_coupling_2012, shalaginov_enhancement_2015, huck_coupling_2016}. 
Despite progress with \SiV{} centers in NDs demonstrating nearly-lifetime limited linewidth \cite{jantzen_nanodiamonds_2016}, they still suffer from surface effects and from a highly strained diamond host with broad inhomogeneous linewidth, large spectral instability and, furthermore, no deterministic way to obtain single NDs containing single \SiV{} centers.

Here we show that \SiV{} centers contained in NDs have improved optical properties after surface treatment in hydrogen plasma, and demonstrate for the first time that ND \SiV{} centers are capable of providing atom-like spectral properties as in bulk diamond.
Single \SiV{} centers are isolated in NDs and their optical transition fine-structure is used to explore the strain of their host crystal environments, with 70\% of the \SiV{} centers exhibiting transversal strain splitting similar to the lowest strain bulk diamond.
The natural distribution of strain in the NDs provides the first systematic inference of the \SiV{} zero-strain orbital splitting, and also the ratio of transverse strain splitting coefficients between ground and excited states.
Low-strain \SiV{} in NDs are shown to exhibit a high degree of optical polarization contrast, almost Fourier-limited linewidth, and spectral stability within the Fourier-limited linewidth over ten minutes.
In addition we demonstrate nanomanipulation abilities necessary to construct arrays of \SiV{} emitters for hybrid photonic platforms. 
An atomic-force microscope probe is used to position NDs in one- and two-dimensional arrays with an accuracy below the uncertainties of the emitter localization of typically $\approx 25$\,nm limited by the extend of the ND and the lateral resolution of the AFM. 
This technique is also shown to be capable of rotating NDs for dipole alignment, and decomposing ND clusters. 
These abilities to control and nanomanipulate single \SiV{} center in NDs with sizes of a few tens of nm and large spectral homogeneity expand a toolbox already consisting of pick-and-place techniques \cite{huck_controlled_2011,wolters_coupling_2012,huck_coupling_2016}.
Our results open new possibilities to optically couple arrays of SiV centers in NDs, exploring novel routes for hybrid quantum technology in fields like quantum networks, quantum sensing, and quantum simulations.

\section*{Results}

\subsection*{Surface treatment of NDs to alter \SiV{} optical properties}

Intrinsic \SiV{} centers form during growth of NDs synthesized under chemical vapor deposition \cite{neu_narrowband_2011} and also under direct high-pressure high-temperature (HPHT) synthesis \cite{davydov_production_2014}.
HPHT conditions effectively allow crystal annealing during growth, which can lead to the production of low-strain NDs.
HPHT NDs have previously been found to host \SiV{} centers with excellent optical properties \cite{jantzen_nanodiamonds_2016}, and these same ND samples were used here.
The NDs were size-selected using a centrifuge (procedure outlined in the methods section) and were spin-coated on IIa diamond substrates containing markers produced by selective carbon milling \cite{jantzen_nanodiamonds_2016}.  
The typical ND size was around or below $130$\,nm, and physical dimensions of specific NDs under study were obtained by SEM imaging that was spatially correlated with fluorescence images using the substrate markers.

\begin{figure}
	\centering
	\includegraphics[width=1\linewidth]{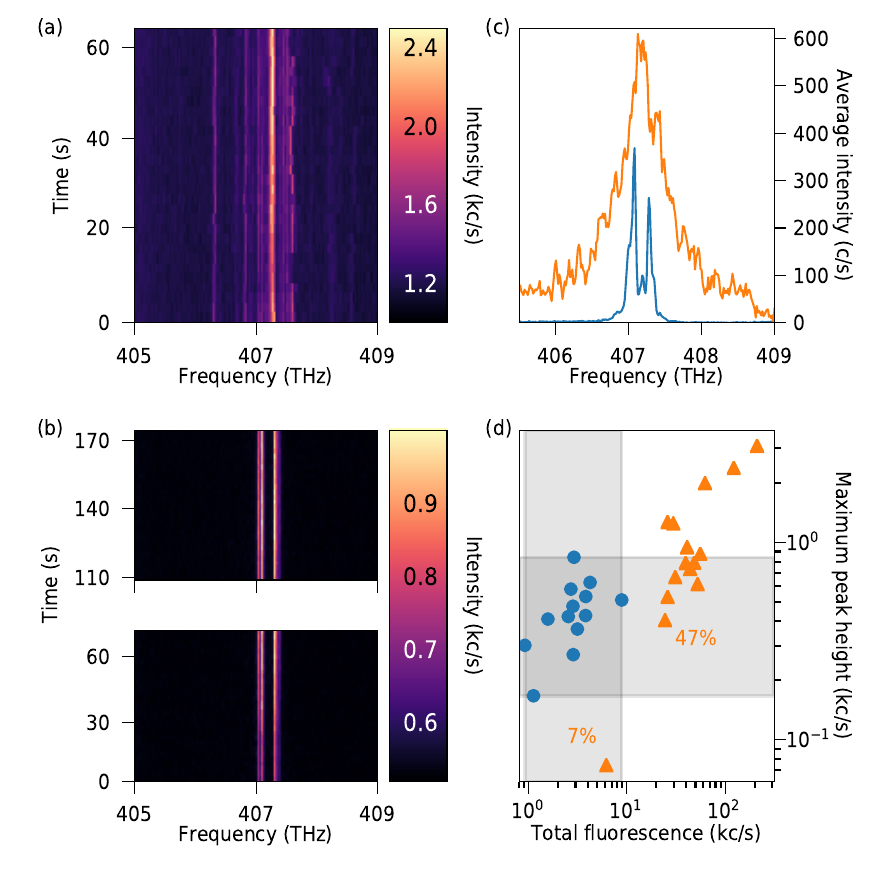}
	\caption{Comparison of optical properties between untreated and H-plasma treated \SiV{}  in NDs at cryogenic temperature. 
		\subfigcap{a} Time-resolved spectrum of \SiV{}  in untreated ND sample exhibiting many lines and spectral diffusion.
		\subfigcap{b} Time-resolved spectrum of \SiV{}  in H-plasma treated ND sample, which presents the 4-line fine structure of \SiV{}  ZPL, with no spectrometer-resolvable spectral diffusion. The discontinuity of time between 70\,s\,$-$\,110\,s was due to refocusing the confocal microscope. 
		\subfigcap{c} The broader orange peak  (15 spots) shows the inhomogeneous spectrum from untreated ND samples. 
		The narrower blue spectrum, where doublet-like fine-structures can be resolved, shows the inhomogeneous spectrum of spectrally-stable points from H-plasma treated ND samples (23 spots). 
		\subfigcap{d} Comparison of the total fluorescence and maximum peak height for spots in the H-plasma treated (blue dots) and untreated (orange triangles) samples. 
		The shaded regions illustrate that the H-plasma treatment reduced the total fluorescence much more than the height of individual peaks in the spectra.
		Only 7\% of spots from the untreated sample had a total fluorescence overlapping with the treated sample spots, while 47\% of the points from the untreated sample had a maximum peak height in the same range as the H-plasma treated sample.
		 } 
	\label{fig:figure2}
\end{figure}

Previous spectroscopic investigations of \SiV{} centers in these HPHT NDs suggested that the optical properties were impacted and in some ways limited by effects on the ND surface \cite{jantzen_nanodiamonds_2016}.
In particular, the photoluminescence (PL) spectrum of the zero-phonon lines of \SiV{} centers exhibit undesired spectral diffusion.
This was characterized with off-resonant excitation at $532$\,nm and at cryogenic temperatures of $8$ K (apparatus described in Methods), where groups of narrow lines around $737$\,nm were observed in photoluminescence (PL) as shown in shown in \autoref{fig:figure2}(a).
These features in the spectrum arise from the direct optical transitions of \SiV{} centers, and it is apparent that they shift position and sometimes blink over the measurement duration.
Taking the time-averaged emission from these measurement and summing across $15$ different spots produced the small-ensemble PL spectrum shown in red in \autoref{fig:figure2}(c).
The characteristic four-line pattern of the \SiV{} zero-phonon line \cite{rogers_electronic_2014, hepp_electronic_2014} cannot be resolved, indicating that the inhomogeneous spread of \SiV{} centers in this ensemble is broader than the \SI{252}{\giga\hertz} excited state splitting.
This result is consistent with previous reports \cite{jantzen_nanodiamonds_2016}, and is a fundamental limitation to many applications when continuous resonant excitation of individual optical transition is required.

A part of this ND sample was treated in a hydrogen plasma (see methods), and time-resolved PL spectra were measured for 23 spots containing \SiV{}.
The PL spectra of \SiV{} transitions in the treated NDs showed significantly improved spectral stability, and a typical measurement is shown in \autoref{fig:figure2}(b).
In addition, the four line fine-structure of \SiV{} was clearly visible in the PL spectrum for 17 of the 23 points investigated (see Appendix B).
For two points we observed broader spectral peaks in a stable doublet, which may be attributed to a thermal broadening of the peaks with higher temperatures due to small contacting area between those NDs and the substrate. 
The small-ensemble spectrum (blue curve in \autoref{fig:figure2}(c)) integrated for the stable spectra in the treated sample clearly has a doublet structure, indicating that the inhomogeneous distribution is narrower than the 252\,GHz excited state splitting.

We compare the PL spectra from the treated sample with the PL spectra from the untreated sample at cryogenic temperature and with identical experimental setting.
As shown in \autoref{fig:figure2}(d), there was only a minor effect on the maximum peak height by H-termination, as $47$\,\% of the  points from the untreated sample lie in the same range as the H-plasma treated ones.
In contrast, the total fluorescence altered significantly (within the range between $733$\,nm to $741$\,nm) with an overlap quantity of only $7$\,\%. 
This indicates that the surface treatment reduces the brightness of the NDs by ``switching off'' some of the \SiV{} centers, but that the remaining ones are just as bright as they were before treatment.
If more than one emitter is present in a focal spot overlapping emission lines could lead to a change in maximum peak height in a spectrum.
However, the large inhomogeneous spectral distribution in the untreated NDs suggests a small likelihood of overlapping emission lines between \SiV{} centers in any given ND.
In addition, \SiV{} centers in untreated NDs gave a g$^{2}(\tau )$ auto-correlation function with no visible dip at $\tau=0$ (the signature of single quantum emitters) whereas \SiV{} fluorescence from the treated NDs showed clear sub-Poissonian emission with g$^{2}(0)$ down to 0.2 after background correction (see Appendix A).
This strengthens the interpretation that the surface treatment has reduced the number of active \SiV{} centers in the NDs, and indicates the possibility to isolate single, most stable \SiV{} center per ND even under off-resonant excitation.

\begin{figure*}
	\centering
	\includegraphics[width=1\linewidth]{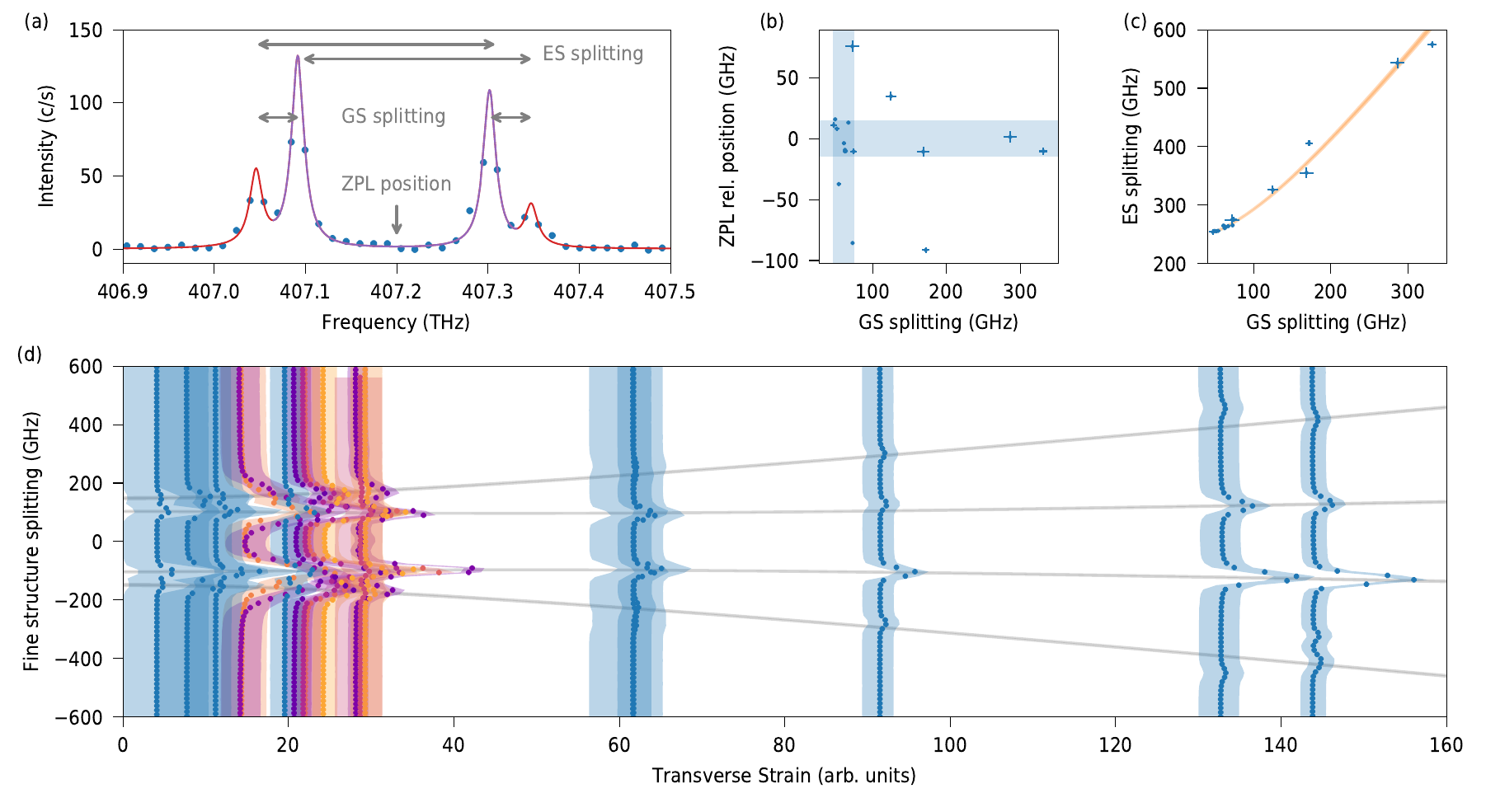}
	\caption{
		Strain analysis of \SiV{} centers in the nanodiamond environment. 
		%
		%
		%
		\subfigcap{a} The four optical transitions A, B, C and D (compare \autoref{fig:array-illustration}) are resolved at low temperature in the PL spectrum. 
		Transverse strain increases the ground (GS) and excited state (ES)  splittings, and axial strain (along with other factors) can alter the ZPL position.
		\subfigcap{b} The four-line PL spectrum was clearly identified in 17 of the 23 spots measured, and the splittings determined.
		There is no correlation between the \SiV{} ground state splitting and the ZPL central frequency, due to the random orientation of strain in the NDs. 
		The $30\times30$\,GHz shaded region is interpreted as low strain and contains 8 out of the 17 NDs. 
		\subfigcap{c} 
		There is close correlation between ground state splitting and excited state splitting.
		The orange region represents the 95\% credible region inferred from a transverse strain model, and these data indicate that the excited state strain splitting coefficient is $1.688^{+0.008}_{-0.008}$ times larger than that of the ground state.
		\subfigcap{d} Statistical inference of the \SiV{} strain splitting from 18 PL spectra (a spot with unresolved four line pattern was suitable for this analysis).
		The grey lines shade the 95\% credible region for the optical transition energies split by transverse strain, and provide the first systematic identification of the \SiV{} zero-field splittings in the ground ($46.3^{+2.5}_{-3.4}$\,GHz) and excited ($252.0^{+1.3}_{-1.8}$\,GHz) states.
		Each spectrum is placed at its inferred transverse strain value, with the shaded band corresponding to the 95\% credible region.
		The spectra between 14 and 30 strain units are colored to aid visual differentiation.
		Thirteen of the \SiV{} spectra correspond to the low-transverse-strain situation where the splitting is dominated by spin-orbit interaction (increasing the chances of spectrally identical emitters).
	}
	\label{fig:strain}
 \end{figure*}

\subsection*{Strain analysis of the fine-structure of \SiV{} in NDs}

Variation in the ZPL lines between NDs is attributed primarily to differences in strain.
The $^2\mathrm{E_g}$ ground and $^2\mathrm{E_u}$ excited states each have twofold orbital degeneracy, and are intrinsically split by spin-orbit interaction as illustrated in \autoref{fig:strain}(a) \cite{rogers_electronic_2014}. 
Strain aligned with the symmetry axis of the \SiV{} center does not lower the symmetry and so cannot lift the orbital degeneracy, but it can shift the energies of the ground and excited states and this appears as a shift in the position of the ZPL.
Strain transverse to the symmetry axis does lift the orbital degeneracy, splitting the $\mathrm{E_x}$ and $\mathrm{E_y}$ orbitals apart in energy.
This is observed as an increase in the splitting between fine-structure peaks in the ZPL, which are resolved in cryogenic PL as shown in \autoref{fig:strain}(b).
The ZPL position and the splittings for ground and excited states were determined for each SiV that showed a clear four-line spectrum.
\autoref{fig:strain}(c) shows the position versus the ground state splitting, which essentially corresponds to axial strain versus transverse strain.
There is no correlation, which is consistent with a random orientation of the strain in the NDs relative to the \SiV{} symmetry axis.
This comparison is further complicated by the numerous unrelated effects such as temperature \cite{jahnke_electronphonon_2015} and isotopic shift \cite{dietrich_isotopically_2014} which can shift the \SiV{} transitions (but not cause splitting).
It is however apparent that almost half of the \SiV{} spectra are clustered fairly close together near the minimum ground-state splitting, shaded in \autoref{fig:strain}(c).
Spectral splitting indicates splitting of the orbital degeneracy in the E ground and excited states.
Unlike the line position, transverse strain is the most likely cause of orbital splitting and so the spectral splitting can provide more information about strain.

The transverse strain was modeled for the four-level system illustrated in \autoref{fig:strain}(a) using the standard strain Hamiltonian
\begin{equation}
\begin{bmatrix}
	\nu + \frac{1}{2}\lambda_\mathrm{SO}^\mathrm{e} & \delta E^\mathrm{e} &  & \\
	\delta E^\mathrm{e} & \nu - \frac{1}{2}\lambda_\mathrm{SO}^\mathrm{e} &  & \\
	& & \frac{1}{2}\lambda_\mathrm{SO}^\mathrm{g} & \delta E^\mathrm{g} \\ 
	& & \delta E^\mathrm{g} & -\frac{1}{2}\lambda_\mathrm{SO}^\mathrm{g}
\end{bmatrix}
\end{equation}
where $\nu$ is the optical transition frequency (ZPL position), $\lambda_\mathrm{SO}^\mathrm{e}$ and $\lambda_\mathrm{SO}^\mathrm{g}$ are the spin-orbit splittings in excited and ground states respectively, and $\delta E^\mathrm{e}$ and $\delta E^\mathrm{g}$ are the strain splitting coefficients in the excited and ground states respectively.
Fitting the acquired four line spectra to our model using Bayesian inference yields a strong correlation  between the excited-state splitting versus the ground-state splitting (\autoref{fig:strain}(d)).
This is expected for transverse strain, which must be the same for ground and excited states of a localized \SiV{} center.
The correlation between ground and excited state splittings in \autoref{fig:strain}(d) gives the strain coefficient ratio $\delta E^\mathrm{e} / \delta E^\mathrm{g} = 1.688^{+0.008}_{-0.008}$ and it is the first time this has been determined for the \SiV{} center.
Only one study has reported uni-axial strain measurements on \SiV{} centers in diamond \cite{sternschulte_luminescence_1994}.

Although the calibration of the strain is unknown in these NDs, the systematic observation of splittings enables a reliable identification of the intrinsic \SiV{} spin-orbit splitting.
Bayesian inference \cite{jaynes_probability_2003, gregory_2005} was used to find the most plausible $\lambda_\mathrm{SO}^\mathrm{e}$, $\lambda_\mathrm{SO}^\mathrm{g}$, $\delta E^\mathrm{e} / \delta E^\mathrm{g}$, and strain values for the \SiV{} spectra.
The use of Bayesian inference allowed the many-dimensional model to be used effectively, and avoided the complications of calculating uncertainty propogation with non-gaussian errors (for more detail see Appendix C).
The results are illustrated in \autoref{fig:strain}(e), and give zero-strain splittings of $\lambda_\mathrm{SO}^\mathrm{g} = 46.3^{+2.5}_{-3.4}$\,GHz and $\lambda_\mathrm{SO}^\mathrm{e} = 252.0^{+1.3}_{-1.8}$\,GHz for the ground and excited states respectively (the full posterior distributions are given in Appendix C.).
These results also indicate that 13 of the 18 \SiV{} spectra are in the ``low transverse strain'' situation, where the splittings are dominated by spin-orbit interaction (below about 40 transverse strain units).
It is remarkable to find such a high proportion of low-strain \SiV{} centers in small NDs.

\subsection*{Recovery of spectral bulk properties}

\begin{figure}
	\centering
	\includegraphics[width=\linewidth]{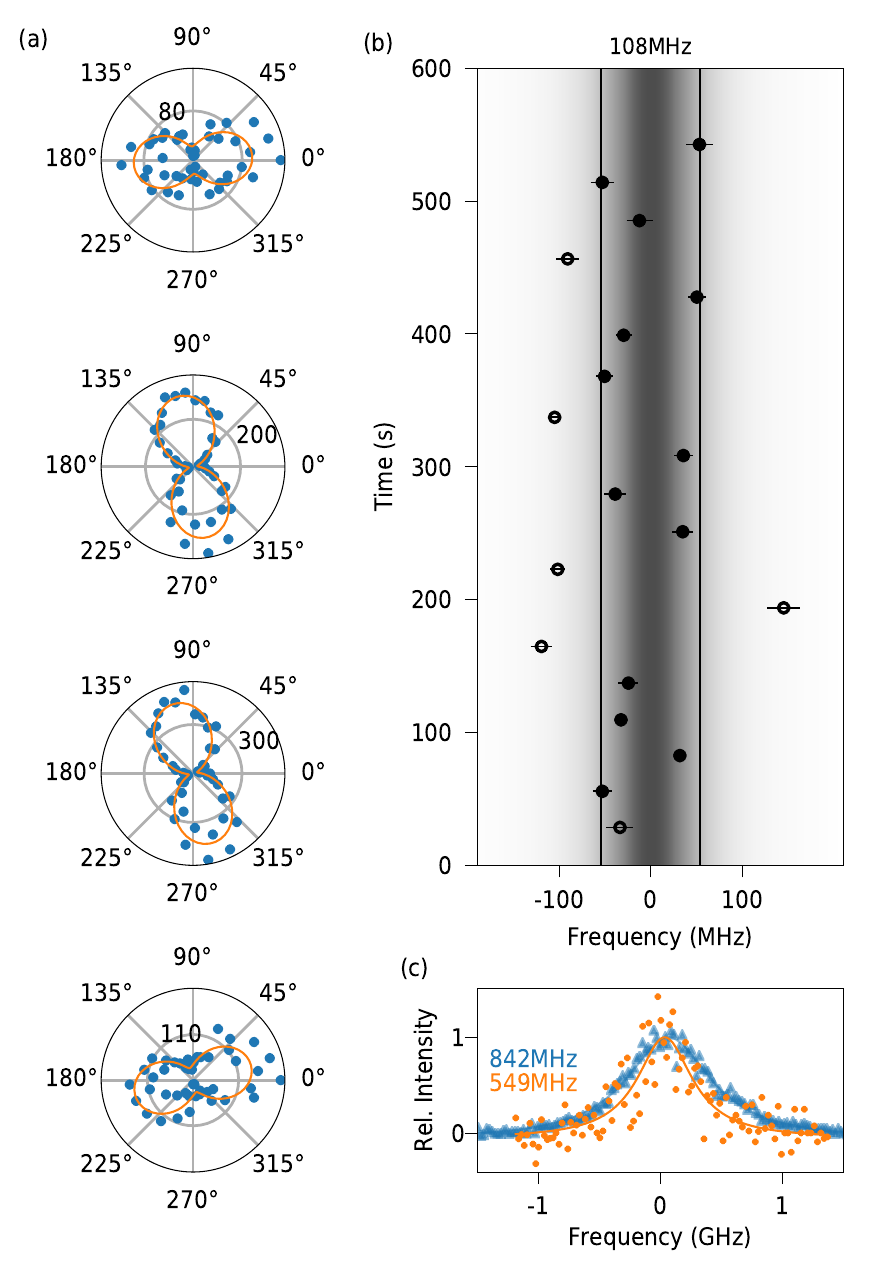}
	\caption{
		Polarization of ZPL and spectral stability under resonant excitation of \SiV{} 	in H-plasma treated NDs. 
		\subfigcap{a} The polarization pattern of the 4-line fine structure  for optical transitions A to D (from top to bottom). 
		The expected behavior of low strain bulk \SiV{} could be recovered. 
		\subfigcap{b} Line position of a resonantly excited \SiV{} center over 600\,s. 
		Occasional blinking destroyed the symmetry of some PLE scans, resulting in ill-fitted peak position (empty dots).
		Excluding such points, the transition frequency remained inside a range of 108\,MHz, as indicted by the shaded band. 
		\subfigcap{c} 
		The averate PLE linewidth (blue triangles) was 842\,MHz and the narrowest scan (orange dots) had a width of 549\,MHz.
		The peak heights were normalized to 1 for better comparison. 
	}
	\label{fig:figure4}
\end{figure}

Resolving the fine-structure also provides the first opportunity to study the polarization behavior of single \SiV{} centers in NDs. 
In the absence of perturbations such as strain, the four optical transitions of the \SiV{} ZPL are known to follow the polarization rules illustrated in \autoref{fig:array-illustration} \cite{rogers_electronic_2014}.
The taller ``inside'' lines B and C arise only from the dipole moment parallel to the \SiV{} axis ($Z$), and have strong polarization contrast.
The outer lines A and D arise from the perpendicular $X, Y$ dipole moments, and viewing geometry means that these always appear ``polarized perpendicular'' to lines B and C (although A and D typically show less polarization contrast because they are not from a single dipole).
Polarization dependence of the spectrum was recorded for many of the \SiV{} sites found in NDs, and a typical example is shown in \autoref{fig:figure4}(a).
The polarization behavior was the same as for \SiV{} in low-strain bulk diamond. 
This adds further support to the interpretation of minimal strain for these \SiV{} centers in NDs, and also demonstrates access to experiments in a cross-polarized configuration. 
%


Related work has shown that \SiV{} in nanodiamonds can have nearly lifetime limited emission \cite{jantzen_nanodiamonds_2016}.
For an indistinguishable photon source it is furthermore important to establish spectral stability within the range of lifetime-limited linewidth for sufficiently long time.
A set of 19 photoluminescence excitation (PLE) scans were recorded sequentially in 600\,s.
In six of these scans the \SiV{} blinked off while the peak was scanned, leading to misfits of the peak position.
The resonance frequencies of the remaining 14 scans were scattered within a frequency band of 108\,MHz.
This PLE measurement is far more precise than the spectrometer-resolved data in \autoref{fig:figure2}.
An excited state lifetime of 1.5\,ns was extracted from $g^{2}(\tau)$ autocorrelation measurements (see Appendix A), corresponding to a lifetime-limited linewidth of $106$\,MHz.
It is concluded that the surface treatment was able to recover excellent spectral stability.

Although the line position was found to be stable within the lifetime-limited linewidth, the PLE linewidths for each of these scans was broader.
The narrowest peak was 549\,MHz, and the data averaged over all scans had a width of 842\,MHz.
This makes it impossible to rule out fast spectral instability on a timescale much shorter than the PLE scan period, but there are other effects that also account for this broadening.
The PLE measurement was made on \SiV{} transition C with an excitation power of 600\,nW at the objective, which is ten times higher than for similar measurements previously perfomed on \SiV{} centres in bulk diamond samples.
It is also difficult to be sure of the temperature of NDs, which have less reliable thermal contact with the cryostat coldfinger than bulk samples typically achieve.
Thermal broadening of the transition would account for the need to use higher laser power in PLE scans, and the use of higher laser power may have caused additional power broadening.

\subsection*{Manipulating NDs on the nanoscale}

\begin{figure} 
	\includegraphics[width=1\columnwidth]{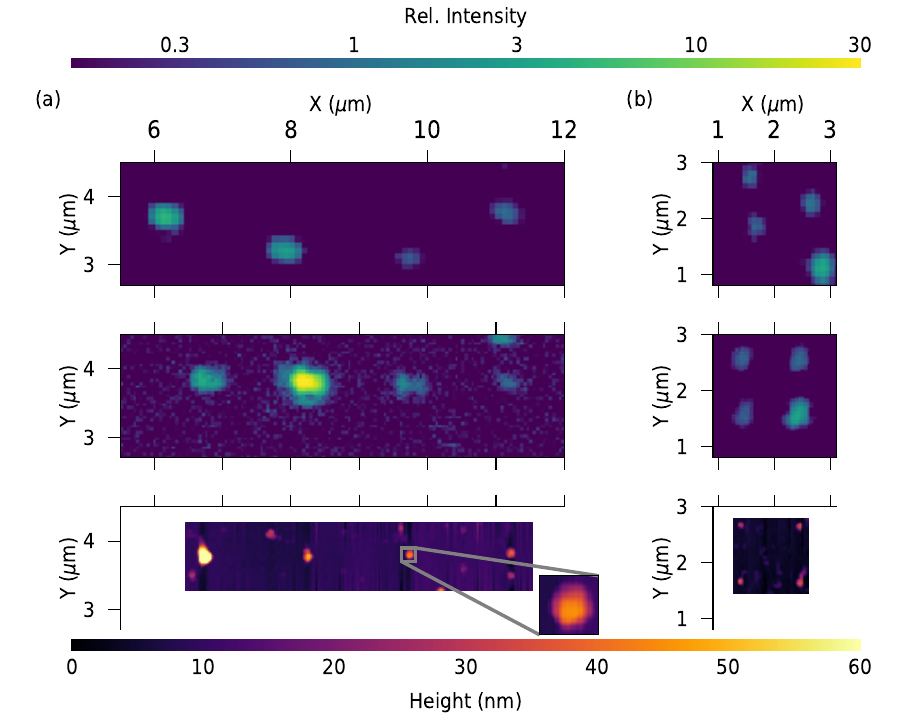}
	\caption{
		Manipulation of NDs at the nanoscale.
		\subfigcap{a} 1D alignment of $4$ NDs with an accuracy of $37$\,nm. 
		From top to bottom: Fluorescence image of spin-coated ND sample on glass; fluorescence image of the same sample area after nanopositioning procedure with the AFM in contact mode; topography image of the aligned 4-ND string with the AC-mode of the AFM. 
		The intensity of fluorescence of emitters were plotted logrithmically, and were normalized by the fluorescence intensity of the encor emitter (far right) for better visual recognition of the spatial positions.
		The inset spans $0.2\times0.2$\,$\mu$m to illustrate the AFM resolution, which has a full-width at half-maximum of 75\,nm.
		\subfigcap{b} 2D nanopositioning with accuracy of $37$\,nm. 
		From top to bottom: Fluorescence image of spin-coated ND sample on glass; fluorescence image after nanopositioning procedure of AFM with contact mode; topography image of the aligned 4-ND square with AC-mode of AFM.  
		The color bars are shared with part \subfigcap{a}.
	}
	\label{fig:nanomanipulation}
\end{figure}

Positioning NDs containing negatively charged Nitrogen-Vacancy (\NV{}) centers in close proximity to plasmonic structures and within the electromagnetic field mode of a photonic crystal cavity has previously been demonstrated using an atomic force microscope (AFM) \cite{huck_controlled_2011,wolters_coupling_2012,huck_coupling_2016}. 
In these studies the positioning precision was determined from the coupling strength to the photonic structure under investigation. 
Here we performed position manipulation of NDs on the
nanoscale with as-yet unreached accuracy using an AFM in high-resolution contact mode and introduce new manipulation tools, such as dipole rotation and ND declustering.
Small NDs with an average diameter of 25\,nm were chosen since the diamond size fundamentally limits the uncertainty of localizing the emitter in a confocal microscope. 
The NDs were preselected to contain \NV{} centers, since these provide a well-established way to spectrally determine alignment angles to an external magnetic field \cite{balasubramanian_nanoscale_2008}.
The \NV{} centers were imaged using an integrated confocal fluorescence microscope.

A string of four NDs was aligned and equally spaced with \SI{1.5}{\micro\meter} separation as shown in \autoref{fig:nanomanipulation}(a).
The confocal fluorescence images show that bright NDs were moved into a straight line configuration.
The AFM lateral resolution is limited by the cantilever tip radius of curvature (nominally $\approx 10$\,nm), and was capable of image spots about 75 nm width (FWHM).
This is considerably better than the diffraction-limited confocal microscope resolution.
The final position coordinates of the NDs were obtained from the center of the two-dimensional Gaussian fits to the AFM topographic data.
The typical deviation of the ND positions from the target coordinates was found to be $\Delta x_{nn}= 37$\,nm and $\Delta y_{nn}= 22$\,nm.

On our device, the pushing operation required a ``field of view'' that included the reference position as well as the desired position.
Measurements of separation distances are less precise for a wider-field scan, and so the ND positioning accuracy diminishes with increasing separation from the reference location. 
ND separations much larger than the optical wavelength were chosen here to extract a lower bound on the positioning accuracy.

Two-dimensional arrays of NDs are more challenging to align, but of great interest for many applications.
This capability was demonstrated by arranging a different set of four NDs into a square pattern with separation of \SI{1}{\micro\meter} as shown in \autoref{fig:nanomanipulation}(b).
The position deviation from the target coordinates was similar to that observed for the straight-line arrangement, but increased slightly for the corner diagonally opposite the reference ND due to accumulation of successive positioning procedures.
All positioning operations were performed with a precision in the order of the ND size, and further precision may be obtained by iterating the procedure with feedback from intermediate measurements of the ND location (see methods for details).
The location of the \NV{} or \SiV{} emitter within a ND is difficult to obtain, and this represents a fundamental limit on the positioning accuracy of the emitter.
We estimate the overall position uncertainty of the emitter is $\approx 40$nm, where the size of the ND is the dominant contributing factor.

\begin{figure}
	\includegraphics[width=1\columnwidth]{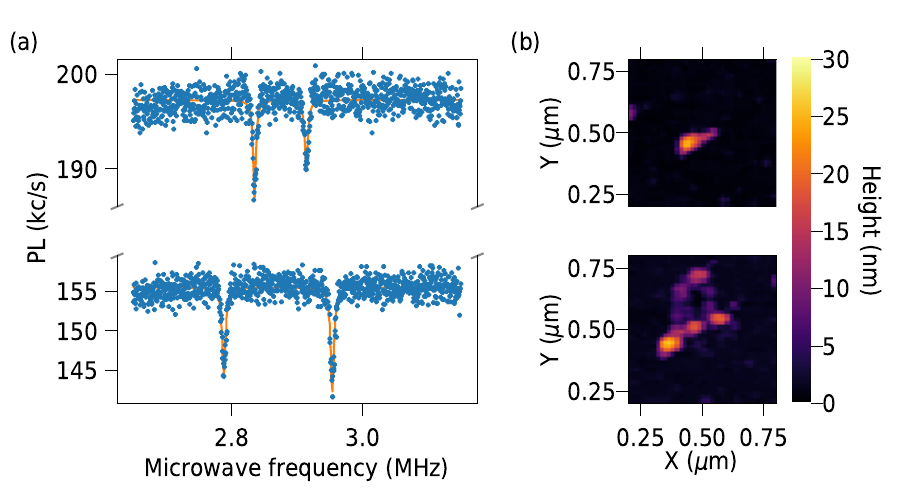}
	\caption{
		In-situ manipulation of NDs.
		\subfigcap{a} Demonstration of NV axis rotation during one tip-push with the AFM in contact mode with pushing distance $\sim 100$\,nm in a constant magnetic field. 
		Before the rotation (above) the resonant frequencies are $2834.8\pm0.1$\,MHz and $2913.8\pm0.2$\,MHz, corresponding to $39\pm4$\,Gauss and the angle between NV axis and magnetic field $\theta = 69\pm2^{\circ}$. 
		After the rotation (below) the resonant frequencies are $2788.3\pm0.1$\,MHz and $2954.0\pm0.1$\,MHz, corresponding to $38\pm4$\,Gauss and an angle $\theta = 38\pm6^{\circ}$. 
		\subfigcap{b} Decomposition of a ND cluster.
		Larger clusters of ND (left image) can be decomposed by pressing the AFM cantilever into the cluster (right image).
	}
	\label{fig:NDrotation}
\end{figure}

Optically-detected magnetic resonance (ODMR) was measured for a single \NV{} center in one ND.
With this measurement information about the alignment of the \NV{} center relative to a stationary external magnetic field from a 3-mm-sized magnet cube, which is sitting about 8\,mm away from the NDs is acquired.
The magnet creates a field of $39\pm4$\,Gauss at the measured ND.
The observed resonance frequencies of $2834.8\pm0.1$ MHz and $2913.8\pm0.2$ MHz (\autoref{fig:NDrotation}(a), uncertainties deduced from Lorentz fitting) correspond to an angle of $69\pm2^{\circ}$ between the \NV{} symmetry axis and the magnetic field.
This ND was moved approximately \SI{100}{nm}, and the ODMR resonance frequencies changed to $2788.3\pm0.1$ MHz and $2954.0\pm0.1$ MHz, corresponding to an angle of a $38\pm6^{\circ}$ relative to the external
field.
It is concluded that the manipulation rotated the ND by approximately $31\pm5^{\circ}$, which is the first demonstration of a potential dipole alignment procedure by rotational ND manipulation. 
The calculation of magnetic field as well as the angle between \NV{} symmetry axis and the magnetic field using ODMR frequencies of \NV{} was done according to Ref. \cite{balasubramanian_nanoscale_2008}.
The comparatively large uncertainties in angle arise mainly from the fact that the zero-field splitting is only known to a precision of $\pm1$\,MHz \cite{acosta2011optical} at room temperature.
For more precise final alignment, an order of magnitude improvement would be possible if the zero field splitting was measured for the specific \NV{} center under investigation (see method section for details).

NDs commonly form clusters when they are dispersed on a substrate, and various techniques have been developed to discourage this phenomenon.
We measured a topological scan of such a ND cluster and were able to show that it is possible to decompose a cluster after it had formed on a substrate by pushing the AFM cantilever into it by applying pressure with the AFM cantilever.
In \autoref{fig:NDrotation}(b) a typical cluster found in our sample is shown as a topological scan.
After pushing the tip into the sample many new features appear in the scan with sizes that are expected for single crystals. 
It is apparent that manipulation with the AFM tip is capable of decomposing a ND cluster into its individual components.

\section*{Discussion}

While it is significant that the H-plasma treatment has altered the spectral properties of \SiV{} centers in these NDs, the mechanism for this change is less obvious.
For \NV{} centers it has been suggested that the strong band bending leads to holes being created in the valence band by H-plasma surface termination and absorption of $\text{H}_2 \text{O}$ \cite{hauf_chemical_2011}. 
Such discharge of the emitters close to the surface for plasma treated NDs containing \SiV{} centers could lead to a reduction of the overall fluorescence as observed and, consequently, only emitters  with larger distance to the diamond surface would remain active.
These are likely to be the least strained \SiV{} centers in the NDs, and so this explanation does account for the reduction in inhomogeneous ensemble linewidth and the high incidence of low-strain spectra.
It is also possible that the more highly strained \SiV{} centers are more susceptible to the discharging effect of the surface termination.
%
%

The reduction or elimination of spectral diffusion is significant, because part of the interest in \SiV{} centers arose from the absence of spectral diffusion in low-strain bulk diamond \cite{rogers_multiple_2014}.
Until now the desirable atom-like spectral properties were only reported for \SiV{}  centers in bulk diamond, although recently they were also seen for \SiV{} centers in nanostructured diamond waveguides \cite{evans_narrow-linewidth_2016} leading to entanglement of two \SiV{} centers \cite{sipahigil_integrated_2016}. 
The degradation of optical properties becomes in particular problematic for \SiV{} centers in small NDs with diameter below $100$\,nm which are desired for hybrid approaches due to low light scattering and precise positioning capabilities.  
Spectral diffusion is a particular problem for the application in arrays of emitters, since inhomogeneity in spectral properties limits the optical coupling between nodes in the array.
This has been a substantial problem for arrays of \NV{} centers created through ion implantation and subsequent annealing and magnetic coupling has been demonstrated \cite{meijer_generation_2005,rabeau_implantation_2006}.
%

%

%
While it is possible to prepare NDs containing both \NV{} and \SiV{} centers an alternative nano-manipulation protocol could as well be realized for NDs containing only \SiV{} centers. 
Only the dipole alignment would require an alternative method.
The high optical polarization contrast shown in \autoref{fig:figure4}(a) could be iteratively measured to achieve the desired rotational orientation without requiring \NV{} centers and external magnetic fields.
In the development of \SiV{} arrays, such rotational manipulation is important to allow the alignment of optical dipoles for optimum coupling.

Since the position of NDs is determined from 2-D Gaussian fits, the uncertainties will increase with ND size.
This can be partially solved by intermediate fitting steps, with the penalty of decreased speed of the manipulation process. 
At the end the most important position accuracy is the location of the quantum emitter, which is then only limited by the size of the NDs. 
For improved super-resolution of the emitter position beyond the size of the ND other techniques, such as stimulated emission depletion microscopy could be utilized \cite{klar_subdiffraction_1999}.  

\section*{Conclusion}

In this work we experimentally demonstrated how to recover symmetry-protected optical transitions of \SiV{} centers in low-strained NDs.
Together with manipulation capabilities of NDs on the nanometer-scale our work opens new routes for bottom-up assemblies of indistinguishable quantum emitter arrays and nanophotonic devices with deterministic coupling and emitter position control much better than diffraction limit.
Individual NDs with size around $100$ nm in diameter contain single \SiV{} centers with bulk-like spectral stability.
Such small NDs are most suitable for hybrid technology due to minimized light scattering and highest positioning precision.
Furthermore, a diamond host with dimensions below $100$ nm introduces a cut-off frequency in the phonon density-of-states beyond $50$ GHz.
The spin coherence time of \SiV{} centers in bulk diamond is limited to $\approx 100$\,ns at $4$\,K due to interaction with acoustic phonons at a frequency corresponding to the ground state splitting of $\approx 50$ GHz \cite{rogers_all-optical_2014, pingault_all-optical_2014, jahnke_electronphonon_2015, becker_ultrafast_2016, pingault_coherent_2017}. 
Our work enables therefore new investigation of prolonged spin coherence times of \SiV{} center in NDs at moderate temperatures. 
So far an extended spin coherence time $T_2$ of $13$\,ms and spin relaxation time $T_1$ exceeding $1$\,s in bulk were demonstrated at temperatures of $100$\,mK in dilution refrigerator \cite{sukachev_silicon-vacancy_2017}. 

Furthermore the measurement of the zero field strain splitting can be used to find the relative transverse strain values  for \SiV{} both  in bulk and nanodiamonds and  as a gauge for optomechanical experiments with \SiV{} centers \cite{meesala_strain_2018}.

\section*{Methods}

\begin{footnotesize}

\subsection*{HPHT-Synthesis of NDs}

The Si doped NDs are synthesis following the receipe in reference \cite{davydov_production_2014}. 
High pressure high temperature synthesis of luminescent NDs was performed on the basis of fluoro-hydrocarbon growth system without catalyst metals \cite{davydov_production_2014}.
Naphthalene C$_{10}$H$_8$ (Chemapol) and highly fluorinated graphite CF$_{1.1}$ (Aldrich Chemical) with the impurity content less than 0.5\,\% were used as initial hydrocarbon and fluorocarbon components of the growth system. 
Tetrakis(trimethylsilyl)silane C$_{12}$H$_{36}$Si$_5$ (Stream Chemicals Co.) was taken as the silicon doping compound. Ternary homogeneous mixture C$_{10}$H$_8$-CF$_{1.1}$-C$_{12}$H$_{36}$Si$_5$ with a Si/C atomic ratio of 0.07 were used as the initial material for diamond synthesis. 
The tablets (5\,mm diameter and 4\,mm height) of cold-pressing starting mixture were placed into graphite container which simultaneously served as a heater of the high pressure “Toroid”-type apparatus. 
The experimental procedure consisted of loading the high-pressure apparatus to 8\,GPa at room temperature, then heating the sample to the desired temperature ($\sim$\,1200\,$^\circ$C), followed by isothermal exposure of the sample during 20\,s at this temperature. The obtained high-pressure states have been isolated by quenching to room temperature under pressure and then complete pressure release. 
The recovered diamond materials have been characterized under ambient conditions by using X-ray diffraction, Raman spectroscopy, scanning and transmission electron microscopes (SEM and TEM).    

\subsection*{Size selection of NDs}
After production the NDs were not size selected and ranged from nm size to a few micrometer size.
In order to size select them centrifugation was employed.
The NDs, which were in a solution of ethanol and micro water were dispersed in 1\,ml of micro water which was then centrifuged at 2000\,rpm and the liquid at the top was transferred to a new container. 
The removed liquid contained the smallest NDs of the original mixture, the one used for the experiments described here.\\
The remaining NDs in the condensed liquid were diluted again with 1\,ml of micro water and the process of dilution and centrifugation was repeated another 3 times for 1000\,rpm, 500\,rpm and 300\,rpm to retrieve solutions with different size of NDs.
As the ND concentration of each liquid with the now size selected NDs was low after this process all new solutions were centrifuged at 5000\,rpm and the liquid at the top mixed of each solution removed and stored in a separate container. 
The concentrated solution that was acquired with centrifugation at 2000\,rpm carries the smallest NDs while the majority of which are around or smaller than 130\,nm when observed by SEM on the same IIa diamond substrate as used in cryogenic measurements. 
Apart from the majority, the extremums of ND size, when observed by SEM, were 30\,nm and 280\,nm. The resolving power of microscopy was constrained by the surface charging of non-electroconductive diamonds substrate, leading to poor image quality for NDs smaller than 30\,nm. 
As for larger sized anomalies, over many SEM sessions only 3 NDs above 200\,nm size were found.
\newline
\label{size selection of NDs}

\subsection*{H plasma treatment}

The surface termination is carried out in a plasma assisted chemical vapor deposition (PACVD) reactor. 
The microwave cavity is formed by a glass cylinder in which the microwave antenna is placed and a movable sample plate, on which the NDs on a single crystal IIa type diamond are mounted.
The chamber is pumped to a vacuum of 10$^{-10}$\,mbar before the process is started. 
Hydrogen at a flow rate of 300\,sccm is pumped into the reactor whereby the pressure is increased step wise to 25\,mbar over 480 seconds. 
In the same time the MW power and temperature are increased, also step wise, to 2.5\,kW and 750$^\circ$C, respectively. 
To stabilize the plasma, the resonator height is also adjusted and the microwave back reflection is set to a minimum, by optimizing the microwave conductor manually. 
These conditions are held for 60 seconds before all parameters are  decreased stepwise to zero in 390 seconds.
During this whole time the NDs are exposed to the plasma which results in a hydrogen surface termination.

\subsection*{Cryogenic confocal microscopy and Sample preparation}

For lowest background fluorescence and best thermal conductivity, a type-IIa diamond was chosen as substrate material ($2\times2$\,mm). 
Markers were etched into the surface with selective carbon milling by Helios 3D FEG.
Substrates were cleaned in a boiling solution ($\approx$ 100\,$^\circ$C) of HC$_l$O$_4$:H$_2$SO$_4$:HNO$_3$ (tri-acid mix).
The untreated ND solution was diluted by chloroform and placed inside an ultra sonic bath for $10$\,min.
$2$\,\textmu l of the diluted sample was spin coated onto the substrate at $5000$ rpm for 40 seconds.

The sample was mounted on the cold-finger of a continuous flow helium cryostat, which was cooled to $8$\,K for imaging with a home-built confocal microscope. 
Experimental control was provided by the Qudi software suite \cite{binder_qudi:_2017}.
The \SiV{} centers were off resonantly excited by $532$\,nm green laser with $120\pm 10$\,\textmu W in front of the objective lens.
Spectra were measured after a $560$\,nm long pass filter, and polarization dependence was measured by rotating the fluorescence using a half-wave plate before recording spectra through a fixed Glan-Thompson polarizer.
This prevents any polarization dependence of the detection instruments from impacting the results.
Statistical processing of spectra was performed using Bayesian inference (see Appendix C for details), and Python packages emcee \cite{foreman-mackey_emcee_2013}, Pymc and Pymc3 \cite{patil_pymc:_2010} were used to perform the Markov-chain Monte Carlo ensemble sampler computations.
The resonant excitation and lifetime stability measurements were performed for power broadened lines to increase the signal to noise ratio thereby increasing the reliability of the data.

\subsection*{AFM positioning procedure}

Prior to positioning, the ND region was imaged in the AFM and a reference ND was selected. 
The positions of other NDs were quickly estimated from a visual inspection of the topographic image, and a pushing distance was determined.
The ND was moved using the AFM tip in contact mode, and the exact mechanics of this process depend on the shape of the AFM tip and the ND and the nature of their contact. 
It is possible for the ND to deviate sideways in undesired directions over the course of the movement, and it is important to avoid pushing the ND over the desired location.
The positioning accuracy may be improved by iterating further cycles of position measurement and fine positioning as well as two-dimensional Gaussian fitting (see below) the position of the ND at every step, however this has not been done here for two reasons.
In the first place, the single-shot positioning accuracy is sufficient for many applications relying on optical coupling.
Secondly, this single-shot procedure gives a useful lower bound on the nanomanipulation accuracy and is a valuable reference for further comparison.

After nanomanipulation, the ND positions were obtained with more precision by fitting each ND spot with a circular two-dimensional Gaussian function.
The full-width at half-maximum was found to be about 75\,nm, but the excellent height resolution of the AFM made it possible to locate the Gaussian fit within about 1\,nm.
Additional uncertainty arises from the fact that  NDs are not circularly symmetric (leading to an effective offset of the Gaussian fit). 
However, these uncertainties are considered to be much smaller than the size of the ND, and so their contribution to the emitter position uncertainty is minimal.

The uncertainties in the calculated magnitude and angle of the magnetic field are a direct result of the 1\,MHz uncertainty in the zero field splitting $D$.
This error is less significant for changes in angles or magnetic fields than for the absolute values, as the D field values does not change for an NV in between measurements.
The uncertainties of the angle and B-field were calculated using Gaussian error propagation with the package \cite{uncertanties_Lebigot}.

\section*{Appendix A: $g^{2}(\tau )$ auto-correlation function measurement}
\begin{figure}[h]
	\centering
	\includegraphics[width=\linewidth]{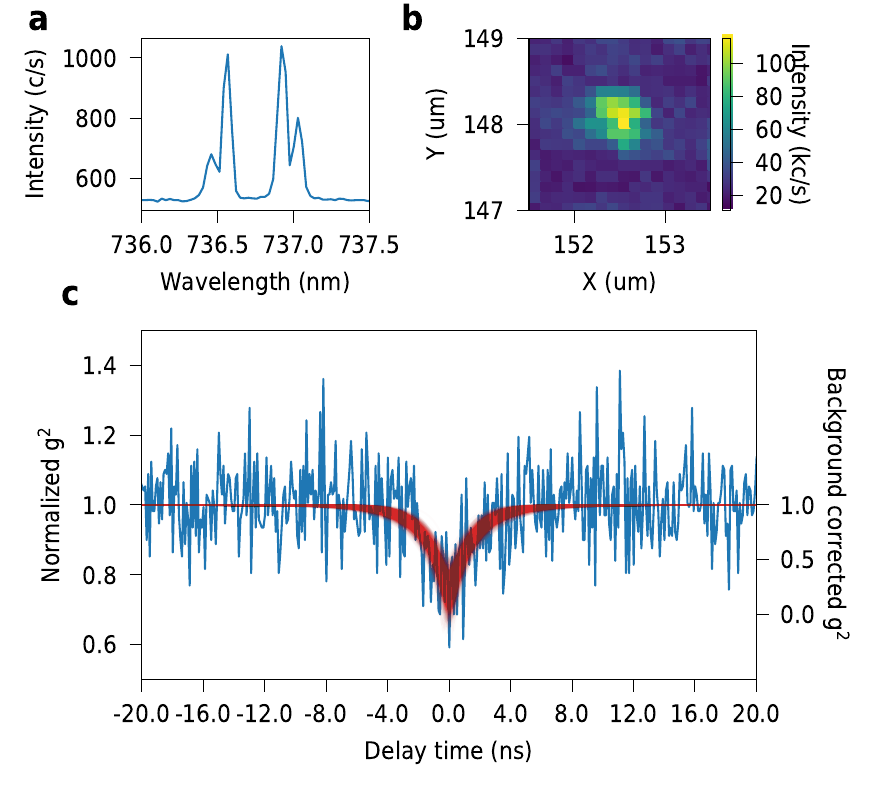}
	\caption{$g^{2}(\tau )$ auto-correlation function measurement shows the existence of single SiV$^{-}$ in ND:
		\subfigcap{a} PL spectrum of the measured single SiV$^{-}$ site.	
		\subfigcap{b} Confocal microscopy image of the measured SiV$^{-}$ site.	
		\subfigcap{c} Normalized g$^{2}(\tau )$ auto-correlation function.
		Measurement data is shown in blue while 95\% credible region of the Bayesian-inference are plotted as a red shading.
		Y axis on the left shows normalized value without back ground correction, while the additional Y axis on the right shows the value of g$^{2}(\tau )$ after background correction.
	}
	\label{fig:figure5}
\end{figure}

$g^{2}(\tau )$ auto-correlation function measurement were conducted to confirm the existence of single SiV$^{-}$ emitters in ND and extract the life time of the excited state.
Measurements were preformed in a Hanbury-Brown-Twiss setup at cryogenic temperatures (8\,K) with off resonant excitation.

\autoref{fig:figure5}c shows a $g^{2}(\tau )$ auto-correlation function measured on a SiV$^{-}$ containing ND, whose Photoluminescence (PL) spectrum (see \autoref{fig:figure5}a) shows typical 4-line pattern Zero-phonon-line (ZPL) at cryogenic temperature.

The measured $g^{2}(\tau )$ data was fitted to a 2-level model, given by
$$g^{2}(\tau ) = 1-ae^{-\frac{\left| \tau \right|}{\tau_{1}}}$$
where $\tau$ is the delay time and $\tau_{1}$ is the life time of the excited state.
The life time of the excited state was extracted as $1.5^{+0.6}_{-0.5}$ ns (95\% credibility).
Following the method of background correction from \cite{brouri_photon_2000}, we estimated the signal-to-background ratio to be 0.55 by fitting the density plot of confocal microscopy scan in \autoref{fig:figure5}b to a 2D Gaussian peak. 
The background subtracted $g^{2}(\tau )$ shows a dip below 0.5, as an indication of a single emitter.

\section*{Appendix B: Post selection of measured photo luminescence spectra}

For the untreated sample 15 photo luminescence (PL) spectra were acquired. 
All of them were used for Figure 2 (one for 2a, all the 15 spots for 2c and 2d) in the manuscript.
\begin{table}[h]
	\caption{List of spots with measured PL spectra of the treated sample and their contribution to figures in the manuscript}
	\begin{tabular}{|c|p{4.cm}|p{3.2cm}|}
		\hline 
		Spot & Contribution & Description \\ 
		number & & \\\hline 
		1 & Fig. 2cd, Fig. 3cde & clear 4-line  \\ 
		\hline 
		2 & Fig. 2cd & multi-site   \\ 
		\hline 
		3 & Fig. 2cd & multi-site \\ 
		\hline 
		4 & Fig. 2cd, Fig. 3cde & clear 4-line \\ 
		\hline 
		5 & Fig. 2cd, Fig. 3e & doublet, single site \\ 
		\hline 
		6 & Fig. 2bcd, Fig. 3cde & clear 4-line \\ 
		\hline 
		7 & Fig. 2cd & doublet, small features around doublet \\ 
		\hline 
		8 & Fig. 2cd, Fig. 3cde & clear 4-line \\ 
		\hline 
		9 & Fig. 2cd, Fig. 3cde & clear 4-line \\ 
		\hline 
		10 & Fig. 2cd, Fig. 3cde & clear 4-line \\ 
		\hline 
		11 & Fig. 2cd, Fig. 3cde & clear 4-line \\ 
		\hline 
		12 & - & unstable emission,  \newline not clear if SiV \\ 
		\hline 
		13 & Fig. 2cd, Fig. 3cde & clear 4-line \\ 
		\hline 
		14 & Fig. 2cd, Fig. 3cde & clear 4-line \\ 
		\hline 
		15 & Fig. 2c, Fig. 3cde & clear 4-line \\ 
		\hline 
		16 & Fig. 2c, Fig. 3bcde, Fig. 4a & clear 4-line \\ 
		\hline 
		17 & Fig. 2c, Fig. 3cde & clear 4-line \\ 
		\hline 
		18 & Fig. 2c, Fig. 3cde & clear 4-line \\ 
		\hline 
		19 & Fig. 2c, Fig. 3cde & clear 4-line \\ 
		\hline 
		20 & Fig. 2c & multi-site \\ 
		\hline 
		21 & Fig. 2c, Fig. 3cde & clear 4-line \\ 
		\hline 
		22 & Fig. 2c, Fig. 3cde & clear 4-line \\ 
		\hline 
		23 & Fig. 2c, Fig. 3cde & clear 4-line \\ 
		\hline 
	\end{tabular} 
	\label{table:table1}
\end{table}
Meanwhile, PL spectra of 23 spots from the H plasma treated sample were measured and according to the different spectral behavior, subfractions of such spots were employed for different analysis.
To clarify the usage of measured points from the H plasma treated sample in different parts of analysis in our paper, as well as for a better understanding of the quality of color centers in the NDs, here we list out all the 23 spots in Table \ref{table:table1} together with brief descriptions on how they contribute to different evaluations.

The requirements for the different analysis were as follows:

\begin{flushleft}
	\begin{enumerate}
		\item{Fig. 2c), treated sample: Stable emission (22 out of 23) of measured spots.}
		\item{Fig. 2d), comparison for optical properties between treated and untreated sample: 
			Identical experiment settings for the PL spectra of the treated and untreated sample. This means only spectra that were not recorded for polarization measurement (no Half-Wave plate and Glan-Thompson in front of the spectrometer) could be compared. 13 out of 23 spots from the treated sample and 15 spots (all) from the untreated sample were used.}
		\item{Fig. 3c), correlation between GS splitting and ZPL position: PL spectra with clear 4 line structure (17 out of 23).}
		\item{Fig. 3d), correlation between GS and ES splitting: PL spectra of single sites with clear 4 line structure (17 out of 23). } 
		\item{Fig. 3e), transverse strain analysis: Single-site spectra (18 out of 23). 
		}
	\end{enumerate}
\end{flushleft}

\section*{Appendix C: Constructing of Strain model and Bayesian inference}

\subsection*{Model}

The positions of the spectral peaks for the SiV fine structure are given by the energy separations between the excited and ground state branches.
\begin{figure*}
	\includegraphics[width=\textwidth]{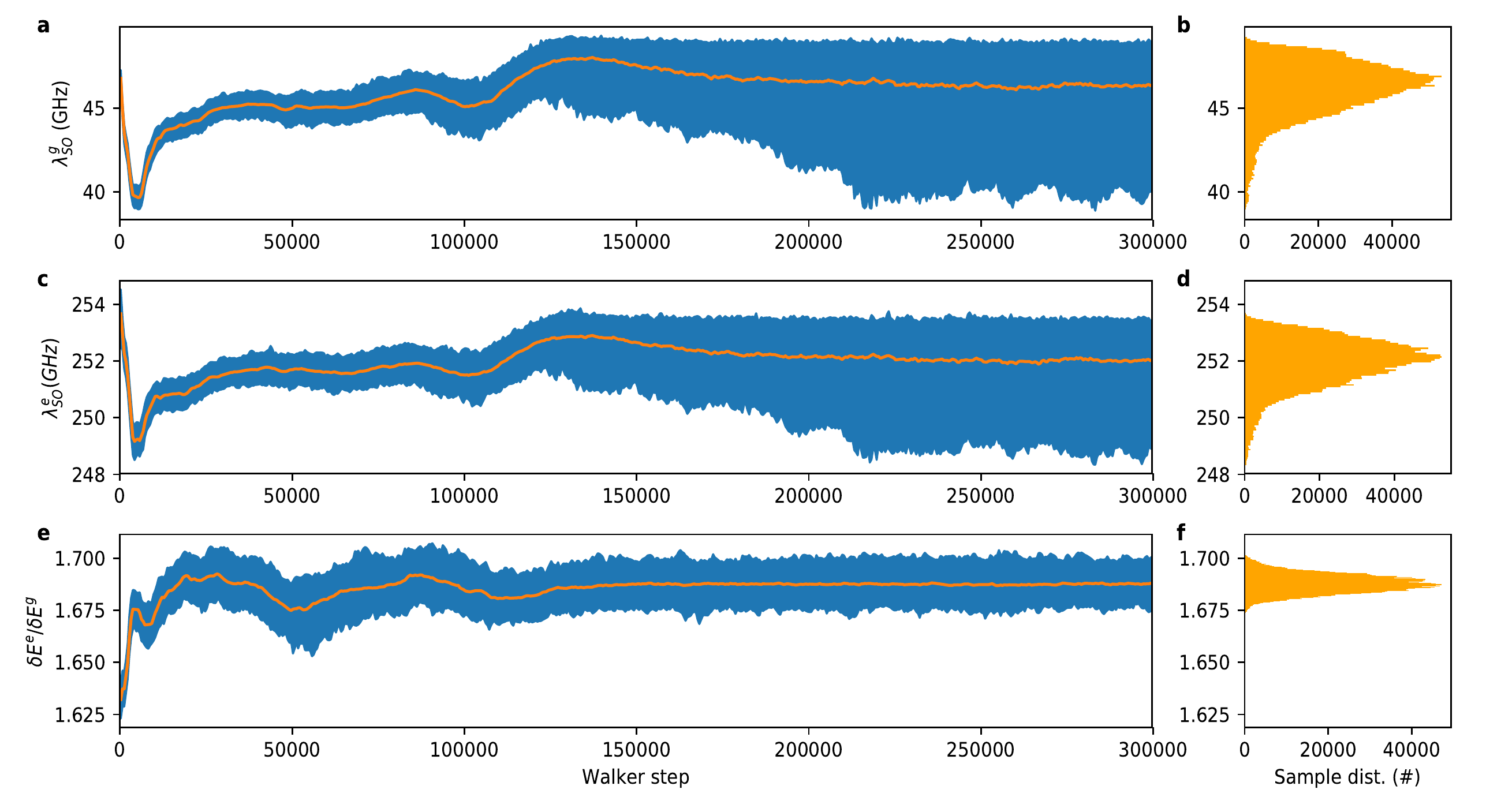}
	\caption{
		MCMC sampler chain trajectories for the fundamental SiV Hamiltonian parameters.
		\subfigcap{a} Median (orange line) and 95\% credible region (blue shading) of the distribution of MCMC walkers for the ground-state spin-orbit splitting $\lambda_\mathrm{SO}^\mathrm{g}$. 
		The credible region and median were calculated for blocks of 50 steps at a time.
		Initial guesses were distributed randomly in the neighbourhood of 47\,GHz.
		On the basis of previously reported SiV ground-state splittings less than 50\,GHz, this parameter was constrained by a prior limiting it to values less than 50\,GHz. 
		It is apparent that from about 125\,000 steps the distribution is skewed by this constraint.
		\subfigcap{b} Histogram showing the final distribution of the MCMC samples, calculated for the final 5\,000 steps.
		\subfigcap{c}, \subfigcap{d} Walker trajectory and final distribution of samples for the excited state spin-orbit splitting $\lambda_\mathrm{SO}^\mathrm{e}$.
		\subfigcap{e}, \subfigcap{f} Walker trajectory and final distribution of samples for the ratio of transverse strain coefficients $\delta E^e / \delta E^g$.
	}
	\label{fig:siv-fundamental-params}
\end{figure*}
The four-level Hamiltonian describing the ground and excited states is given by
$$\begin{bmatrix}
\nu + \frac{1}{2}\lambda_\mathrm{SO}^\mathrm{e} + S_\mathrm{A} & \delta E^\mathrm{e} &  & \\
\delta E^\mathrm{e} & \nu - \frac{1}{2}\lambda_\mathrm{SO}^\mathrm{e} + S_\mathrm{A} &  & \\
& & \frac{1}{2}\lambda_\mathrm{SO}^\mathrm{g} & \delta E^\mathrm{g} \\ 
& & \delta E^\mathrm{g} & - \frac{1}{2}\lambda_\mathrm{SO}^\mathrm{g}
\end{bmatrix}
$$

where $\nu$ is the optical transition frequency, $\lambda_\mathrm{SO}^\mathrm{e}$ and $\lambda_\mathrm{SO}^\mathrm{g}$ are the spin-orbit splittings in excited and ground states respectively, and $\delta E^\mathrm{e}$ and $\delta E^\mathrm{g}$ are the strain splittings in the excited and ground states respectively.
In this slightly simplified picture, $S_\mathrm{A}$ is called "axial strain" but it has the effect of shifting the excited states with respect to the ground and not altering any splitting.
This term encompasses all of the indistinguishable physical effects (temperature, axial strain, hydrostatic pressure, etc) that can shift the spectral lines without altering their pattern.

The excited- and ground-state strain splittings $\delta E^\mathrm{e}$ and $\delta E^\mathrm{g}$ arise from the same transverse strain (which must be independent of the excitation state of the SiV center).
However the two electronic states can have different splittings, and so we consider the transverse strain coefficients $c^\mathrm{g}$ and $c^\mathrm{e}$ that dictate the way the system responds to a transverse strain $S_\mathrm{T}$.
It is impossible for us to know the transverse strain coefficients $c^\mathrm{g}$ and $c^\mathrm{e}$ absolutely without having knowledge of the strain magnitude at each spectrum. 
However, we can find the ratio $cR^\mathrm{eg} = c^\mathrm{e} / c^\mathrm{g}$.

The ideal signal in the absence of background or noise would be made up of four Gaussian peaks. 
We take Gaussian line shapes because the spectrometer measurements are instrument limited. For this same reason, the linewidths are the same for each of the four lines, but their amplitudes ($A, B, C, D$) and positions ($\nu_A, \nu_B, \nu_C, \nu_D$) are different.
So we have expected counts $c_\mathrm{e}$ as a function of $\nu$ given by

\begin{align*}
	c_\mathrm{e} = 
	& A\mathrm{e}^{-(\nu-\nu_A)^2 / 2\sigma^2}
	+ B\mathrm{e}^{-(\nu-\nu_B)^2 / 2\sigma^2}\\
	&+ C\mathrm{e}^{-(\nu-\nu_C)^2 / 2\sigma^2}
	+ D\mathrm{e}^{-(\nu-\nu_D)^2 / 2\sigma^2}
\end{align*}

The transition frequencies ($\nu_A, \nu_B, \nu_C, \nu_D$) are determined from the Hamiltonian and so depend on $\lambda_\mathrm{SO}^\mathrm{g}, \lambda_\mathrm{SO}^\mathrm{e}, cR^\mathrm{eg}, S_\mathrm{T}, S_\mathrm{A}$.

Disregarding background for now, this expected profile of counts vs optical frequency is sampled by the signal which is counted as discrete photons. 
The probability of observing a certain signal photon count at a given frequency is the Poissonian sampling distribution around $c_\mathrm{e}$, i.e.
$$ P(c_s | \nu I) = \mathrm{e}^{-c_\mathrm{e}} \frac{{c_\mathrm{e}}^{c_s}}{c_s !} $$
This involves $c_\mathrm{e}$ and inherits a dependency on the parameters of the 4-Gaussian shape, so it is really $P(c_s|\nu ABCD\sigma \lambda_\mathrm{SO}^\mathrm{g} \lambda_\mathrm{SO}^\mathrm{e} cR^\mathrm{eg} S_\mathrm{T} S_\mathrm{A} I)$.
Here $I$ represents all of the knowledge of the system and the experiment that is not otherwise explicitly stated.
To prevent underflow errors in computation, we need to consider the log probability
\begin{align}
	\log P(c_s|\nu ABCD\sigma \lambda_\mathrm{SO}^\mathrm{g} \lambda_\mathrm{SO}^\mathrm{e} cR^\mathrm{eg} S_\mathrm{T} S_\mathrm{A} I) \\= -c_\mathrm{e} + c_s * \log c_\mathrm{e} - \log\left(c_s !\right)
\end{align}

The background counts are from the CCD and are poisson distributed around an average background rate $\beta$. 
The probability of a background count $c_{bg}$ is therefore
$$P(c_{bg}|I) = \mathrm{e}^{-\beta}\frac{\beta^{c_{bg}}}{c_{bg}!}$$
This depends on $\beta$, and so it is really $P(c_{bg} | \beta)$. Calculating this is tricky because $\beta^{c_{bg}}$ and $c_{bg}!$ get huge leading to overflows. It is more tractable to consider 
$$\log P(c_{bg} | \beta I) = -\beta + c_{bg}\log\beta - \log\left(c_{bg}!\right)$$

Now it is possible to consider the actual posterior distribution.
In other words, given a certain set of model parameters, what is the probability of observing a particular count value $c$?
This is slightly complicated, because the observed counts come from the signal and the background, but we don't have any way of knowing the relative contributions.
Instead of answering this, we can give the posterior probability as a sum of possible (and mutually exclusive) options. 
Either we had $c$ signal counts and zero background counts, or $c-1$ signal counts with 1 background count, or so on.

This gives
\begin{align}
	P(c | \nu ABCD&\sigma \lambda_\mathrm{SO}^\mathrm{g} \lambda_\mathrm{SO}^\mathrm{e} cR^\mathrm{eg} S_\mathrm{T} S_\mathrm{A} \beta I)\\
	& =  \sum_{n = 0}^c \mathrm{e}^{-c_\mathrm{e}} \frac{{c_\mathrm{e}}^{n}}{n !} \mathrm{e}^{-\beta}\frac{\beta^{(c-n)}}{(c-n)!}\\
	& = \mathrm{e}^{-(c_\mathrm{e}+\beta)} \sum_{n = 0}^c \frac{{c_\mathrm{e}}^{n}}{n !} \frac{\beta^{(c-n)}}{(c-n)!}\\
	& = \frac{\mathrm{e}^{-(c_\mathrm{e}+\beta)}}{c!} \sum_{n = 0}^c \binom{c}{n} {c_\mathrm{e}}^{n} \beta^{(c-n)} \\
	& = \frac{\mathrm{e}^{-(c_\mathrm{e}+\beta)}}{c!} (c_\mathrm{e} + \beta)^c
\end{align}
\begin{figure*}[t]
  \includegraphics[width = \linewidth]{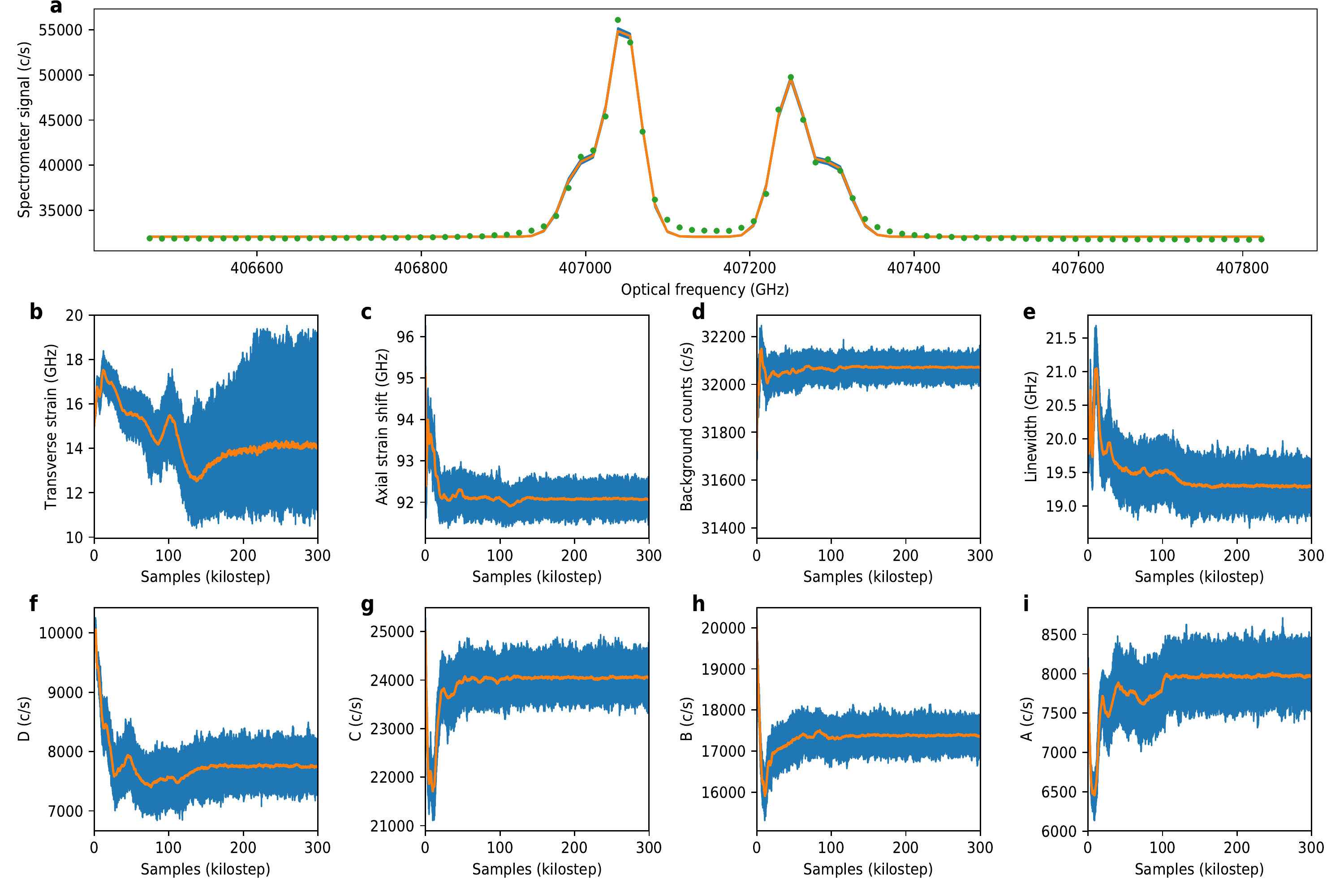}
\caption{
		Spectrum and MCMC sampler chain trajectories for parameters specific to SiV Spot 1.
		\subfigcap{a} The Bayesian analysis infers hamiltonian parameters directly from the raw spectral data (green dots). 
		The final ``fit'' is represented by the median (orange line) and 95\% credible region (blue shading).
		\subfigcap{b} The MCMC sampler chain trajectory for the transverse strain of this spot.
		This parameter provides the horizontal positioning of the spectrum in the strain summary figure (\autoref{fig:siv-fundamental-params} and Figure 3e).
		\subfigcap{c} Sampler trajectory for the axial strain term for this spot. 
		This effects the shift of the ZPL and therefore considers all four peaks together, and so it is not surprising that this parameter reaches equilibrium in the sampler trajectory more quickly than the transverse strain.
		\subfigcap{d} Sampler trajectory for the background count level.
		There are a lot of CCD pixels that measure almost nothing but background, and so it is unsurprising that this parameter quickly reaches equilibrium in the sampler trajectory.
		\subfigcap{e} Sampler trajectory for the linewidth.
		Since linewidth is limited by the instrument or by temperature, our model takes the same width for all four peaks in the spectrum.
		\subfigcap{f} -- \subfigcap{i} Sampler trajectories for the amplitudes D, C, B, A of the four peaks. 
		Peak position matters for the strain analysis, but peak height is not important.
		Here it is visually apparent that sensible values have been obtained.
	}
	\label{fig:mcmc-trajectories-spot1}%
\end{figure*}
The problematic sum has been handled using the binomial sum relation. Again, in order to avoid overflow errors for large $c$, we consider the log probability
\begin{align}
	& \log \left( P(c | \nu ABCD\sigma \lambda_\mathrm{SO}^\mathrm{g} \lambda_\mathrm{SO}^\mathrm{e} cR^\mathrm{eg} S_\mathrm{T} S_\mathrm{A} \beta I) \right) \\
	& = -(c_\mathrm{e} + \beta) - \log(c!) + c\log(c_\mathrm{e} + \beta)
\end{align}
This probability is formally called the ``likelihood''.

\vspace{3mm}
\subsection*{Bayesian inference}

With this encoding of the model into an expression for the likelihood of observed count values at each pixel of each spectrum, we can apply Bayes' Rule to obtain
\begin{align}
	P(ABCD&\sigma \lambda_\mathrm{SO}^\mathrm{g} \lambda_\mathrm{SO}^\mathrm{e} cR^\mathrm{eg} S_\mathrm{T} S_\mathrm{A} \beta | \vec{c} I) \\
	= \sum &P(c | \nu ABCD\sigma \lambda_\mathrm{SO}^\mathrm{g} \lambda_\mathrm{SO}^\mathrm{e} cR^\mathrm{eg} S_\mathrm{T} S_\mathrm{A} \beta I) \\
	&\times P(\nu ABCD\sigma \lambda_\mathrm{SO}^\mathrm{g} \lambda_\mathrm{SO}^\mathrm{e} cR^\mathrm{eg} S_\mathrm{T} S_\mathrm{A} | \beta I) \\
	&/P(c | I)
\end{align}
where the sum is across all pixels of all the spectra under consideration and all but the parameters $\lambda_\mathrm{SO}^\mathrm{g}$, $\lambda_\mathrm{SO}^\mathrm{e}$, and $cR^\mathrm{eg}$ are different between each spectrum.

This is an elaborate computation, because for the 17 spectra examined here there are 139 free parameters to be inferred.
The prior (14) for this inference consisted of various constraints, comprising
\begin{itemize}
	\item{all linewidths $\sigma$ must be positive;}
	\item{all transverse strains $S_\mathrm{T}$ must be positive;}
	\item{the ground-state spin-orbit splitting $\lambda_\mathrm{SO}^\mathrm{g}$ must be less than 50\,GHz (on the basis of published data for SiV with observed ground state splittings less than 50\,GHz);}
	\item{all peak amplitudes $A, B, C, D$ must remain above 60\% of their (manually-obtained by visual approximation) initial guesses.}
\end{itemize}

The probability given in (10) was obtained using a Markov-chain Monte Carlo (MCMC) ensemble sampler algorithm. 
Many of the model parameters (such as peak amplitude) are not important for the discussion of transverse strain, and it is possible to integrate them out by invoking the fact that the integral across all space of any probability must have a value of 1.
In this way it is possible to obtain the probability
$
P(\lambda_\mathrm{SO}^\mathrm{g} \lambda_\mathrm{SO}^\mathrm{e} cR^\mathrm{eg} | \vec{c} I)
$
which represents knowledge of fundamental SiV parameters on the basis of all the PL spectra that are considered.
The MCMC sampler chain trajectories and final probability distributions for these three Hamiltonian parameters are shown in \autoref{fig:siv-fundamental-params}.
\newline
\indent It is important to point out that the probability distributions are not Gaussian, and that especially the $\lambda_\mathrm{SO}^\mathrm{g}$ is not even symmetric.
For the ground state spin-orbit splitting the asymmetry is clearly connected to the prior, which constrains this value to remain under 50\,GHz.
A similar shape is observed in \autoref{fig:siv-fundamental-params}d, for the excited state spin-orbit splitting.
This occurred even though no constraint was imposed on $\lambda_\mathrm{SO}^\mathrm{e}$, and demonstrates the tight relationship established between these parameters on the basis of the spectral data.
Indeed, this is confirmed by the narrow distribution of the splitting coefficient ratio in \autoref{fig:siv-fundamental-params}f.
The non-Gaussian nature of these distributions means that the uncertainties (which have been given as 95\% credible regions) may in places overstate the breadth of the probability distribution if interpreted as relating to standard deviations.

\begin{figure*}[t]
	\includegraphics[width=\textwidth]{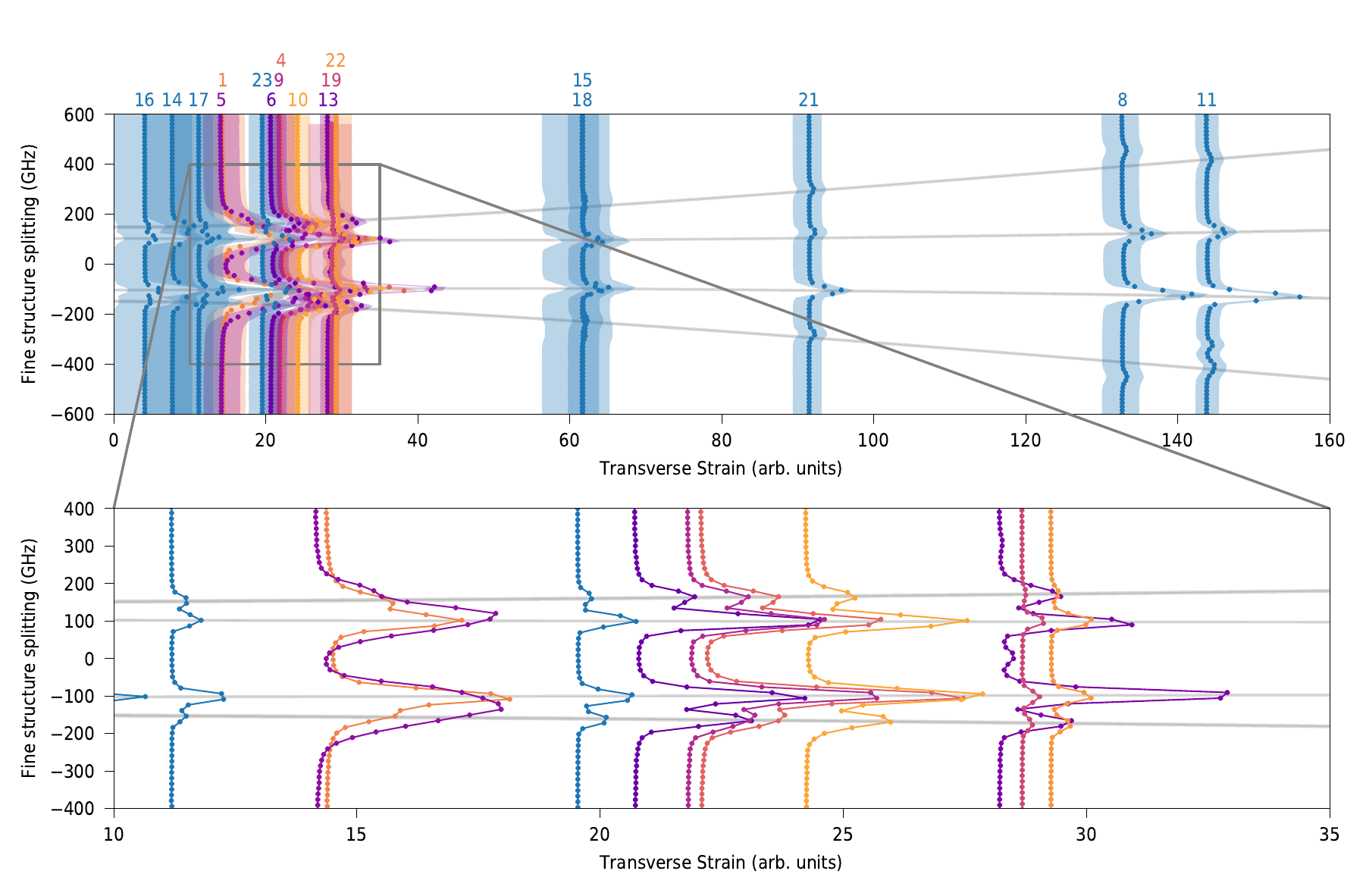}
	\caption{
		Strain analysis with expanded detail. 
		This is a reproduction of Figure 3e in the main manuscript, with the spectra annotated by their ``focal spot number''. 
		Not all of the recorded spectra had a clear enough ``single SiV'' structure to be used in this analysis. 
		The inset shows more detail of the dense region at low strain, which contains numerous almost-overlapping spectra.
		The 95\% credible-region shading has been omitted from the inset for better viewing of the individual spectra.
	}
	\label{fig:strain-inset}
\end{figure*}
\subsection*{Discussion}

Beyond the three fundamental SiV parameters represented in \autoref{fig:siv-fundamental-params}, each PL spectrum has eight additional parameters required to describe it according to the model presented here.
It is possible to examine the MCMC sampler trajectories for all of these parameters, and the spectrum for Spot 1 is shown in \autoref{fig:mcmc-trajectories-spot1} as a representative example.
For each spectrum, suitable ``approximate values'' were obtained for these parameters in a manual visual process. 
The MCMC sampler seed values were randomly generated in a range of a few percent around these visual estimates.
It is clear that some parameters reach their equilibrium distributions quickly, and these tend to be those that are not ``coupled'' to any others (such as the background count level).

Careful examination of \autoref{fig:strain-inset} reveals that spots 8 and 18 appear to be misplaced on transverse strain axis, towards lower strain than the outer peaks A and D would indicate.
While it is normal for fitted data, not be represented exactly by a fit the prior in Bayesian inference favors here the significance of lines B and C and therefore seems to increase increase the error for A and D.
This has occurred due to the outer peaks A and D being much weaker in these spectra than the inner peaks B and C.
As a result, the MCMC sampler has prioritized matching the splitting of inner peaks at the expense of the outer peaks.
To do this the MCMC sampler has reduced its ``fit'' inference for the amplitudes A and D, and has in fact ``hit against'' the minimum peak height constraint of the prior.
In this case, it is mathematically more likely that the small outer peaks arise due to noise than that the inner peaks are slightly offset from their ``true'' position.
Any more extreme mistreatment of the outer peaks A and D was prevented by the prior constraints that were imposed for the amplitudes of these peaks (they were required to stay above 60\% of the manually-identified initial guesses).

The expanded view in \autoref{fig:strain-inset} shows that spot 5 has a linewidth considerably greater than for most of the other spots, and the ground-state splitting is not resolved in the spectrum.
This is interpreted to be the result of elevated temperature, probably occurring due to poor thermal contact between the ND and the cold-finger of the cryostat.
The inability to resolve the ground state splitting means that this spectrum was not included in  Figure 3c and 3d (main manuscript).

Since the spectra were fit using the strain model for the SiV spectrum, spurious peaks in the measurements had little impact.
Spots 11 and 13 clearly have an extra peak, but the model requires symmetric splittings in the spectrum and so any attempted fit to the ``wrong'' peak was automatically suppressed.
This extra peak would have marginally offset the background counts level but not effected any other parameter.

In the strain model, the transition energies remain fairly constant for low strain.
This is due to the splittings being dominated by spin-orbit interaction.
As a result, it is difficult to precisely determine the strain value for a spectrum with minimal strain splitting.
For this reason, the 95\% credible region for transverse strain is expanded for spots 16, 14, and 17.
Spots 16 and 14 had their MCMC sampler distributions for transverse strain ``hit against'' the constraint that required positive transverse strain splitting.
All in all this does not mean, that the strain model misrepresents many points. 
The Bayesian inference algorithms does not fit single spectra but the whole set of data to a model and determines the values for that fit the dataset as a whole. 
Therefore there are not individual fits for an individual spectrum. 
Every line fit is merely a projection of the model onto the transverse strain value

\newpage

\end{footnotesize}


\section*{References}
\bibliography{Reference}

\section*{Acknowledgements}

%
%
We acknowledge G. Neusser and the FIB Center UUlm for etching the markers into the diamond IIa substrate. 
We acknowledge help from F. Frank (Uni Ulm) with the AFM, A. Gilchrist (Macquarie Uni) for guiding discussions in developing the Bayesian analysis, and S.\,\,Strehle (EBS, Uni Ulm) for experimental support.
LR is the recipient of an Australian Research Council Discovery Early Career Award (project number DE170101371) funded by the Australian Government.
VAD thanks the Russian Foundation for Basic Research (Grant No. 18-03-00936) for financial support.
YL is currently supported by Sino-German (CSC-DAAD) Postdoc Scholarship Program (57251553).
ABF acknowledges support of the Carl-Zeiss Foundation.
FJ acknowledges support of the DFG, BMBF, VW Stiftung and EU (ERC,DIADEMS).
AK acknowledges the generous support of the DFG, the Carl-Zeiss Foundation, IQST, the Wissenschaflter-R\"uckkehrprogramm GSO/GZS.

\section*{Author contributions}

Experiments were conceived by LR, AK, and FJ.
OW, ABF, and LR performed the spectroscopic measurements; OW and CO performed the surface treatment; YL and LA performed nanomanipulation of the nanodiamonds. 
ABF and OW did the sample preparation.
ABF preformed the SEM measurements.
LR and OW performed the Bayesian strain analysis.
VAD and VA synthesized the NDs with \SiV{} centers.
The manuscript was written by OW, ABF, LR, and AK and all authors discussed the results and commented on the manuscript.

\section*{Competing financial interests}
The authors declare no competing financial interests.

\end{document}